\documentstyle[11pt,aaspp4]{article}

\slugcomment{Accepted for publication in {\em Ap.\ J.\ Lett.}}

\lefthead{Blundell et al.}
\righthead{Evidence for a black hole in a radio-quiet nucleus}
\begin{document}

\title{Evidence for a black hole in a radio-quiet quasar nucleus}

\author{Katherine M.\ Blundell\altaffilmark{1}, Anthony J.\ 
Beasley\altaffilmark{2}, Mark Lacy\altaffilmark{1} and Simon T.\ 
Garrington\altaffilmark{3}}

\altaffiltext{1}{Astrophysics, University of Oxford, Keble Road, Oxford,
OX1 3RH, U.K.} 
\altaffiltext{2}{National Radio Astronomy Observatory, Soccoro, NM 87801, U.S.A.}
\altaffiltext{3}{University of Manchester, Nuffield Radio Astronomy 
Laboratories, Jodrell Bank, Cheshire, SK11 9DL, U.K.}

\begin{abstract}

We present the first milli-arcsecond resolution radio images of a
radio-quiet quasar, detecting a high brightness temperature core with
data from the VLBA.  On maps made with lower-frequency data from
MERLIN and the VLA jets appear to emanate from the core in opposite
directions, which correspond to radio-emission on arcsecond scales
seen with the VLA at higher frequencies.  These provide strong
evidence for a black-hole--based jet-producing central engine, rather
than a starburst, being responsible for the compact radio emission in
this radio-quiet quasar.

\end{abstract}


\section{Introduction}

The quasar population is divided into two classes: radio-loud quasars
(RLQs) and radio-quiet quasars (RQQs). These two populations are seen
to be distinct in several respects. For example, RLQs have ratios of
total radio luminosity at 5 GHz to optical luminosity $R \approx $ 10
-- 100, whereas RQQs have $R \approx $ 0.1 -- 1 (Kellermann et al.\
1989). Such distinctions are also apparent in the narrow-line
luminosity -- radio luminosity plane (Miller, Rawlings and Saunders
1993) and in plots of far-infrared luminosity {\em vs.}\ radio
luminosity (Sopp and Alexander 1991a).  Furthermore, all RQQs seem to
have luminosities at 5 GHz below $10^{25}\ {\rm W\, Hz^{-1}\,
sr^{-1}}$ (Miller, Peacock and Mead 1990).

The physical reason for this bimodality is not clear; while there is
compelling evidence for a relativistic-jet-producing central engine
(almost certainly involving a black hole) as the source of the radio
emission in RLQs (Begelman, Blandford and Rees 1984), the mechanism by
which the weaker radio emission in RQQs arises is uncertain. It has
been proposed (Sopp and Alexander 1991a) that the radio emission from
RQQs is due to a circumnuclear starburst. In this scenario, radio
emission originates from synchrotron-emitting electrons accelerated in
supernova remnants and/or flat spectrum thermal bremsstrahlung from
HII regions. Indeed, it has been argued that the entire RQQ phenomenon
can be produced by a massive starburst (e.g., Terlevich [1990],
Terlevich and Boyle [1993], Terlevich et al.\ [1995]).

An alternative explanation (Miller et al.\ 1993) is that the radio
emission arises from weak radio-jets originating from the active
galactic nucleus (AGN) in a scaled-down version of the mechanism
present in RLQs. An important test between these alternative
hypotheses is the measurement at high angular resolution of the
brightness temperature ($T_{\rm B}$) and structure of the radio
emission.  If the emission arises in a star-forming region, we might
expect to see the emitting region resolved into a number of small
sources, each with brightness temperature $\stackrel{<}{_{\sim}}10^5
{\rm K}$ (Muxlow et al.\ 1994). If, however, the emission arises from
an AGN, we would expect to see an unresolved point source, and/or a
jet, with a high brightness temperature, and possibly with pc-scale
features having some correspondence to features on the kpc-scale. To
date, only nearby Seyfert galaxies have been the target of very high
resolution radio imaging and the results have been ambiguous, with
$\approx 50$\% showing high ($\stackrel{>}{_{\sim}} 10^6$ K) $T_{\rm
B}$ emission (e.g., Ulvestad et al.\ 1987, Roy et al.\ 1994, Lonsdale
et al.\ 1992). The much lower bolometric luminosities of these objects
compared with RQQs, however, make direct comparisons difficult.

To carry out such a test on a RQQ we used very long baseline
interferometry (VLBI) techniques to examine E1821+643.  This quasar is
radio-quiet (see Lacy et al.\ 1992) with $R \approx 1.5$ and is highly
luminous in all wavebands from the infra-red (Hutchings and Neff 1991)
to the X-ray (Pravdo and Marshall 1984) and since it is at the
moderate redshift of 0.298 (Schneider et al.\ 1992), its radio flux
density is high enough to allow detailed mapping with the Very Large
Array (VLA) telescope. Our VLA images of this object (Blundell and
Lacy 1995) showed that, besides steep spectrum extended emission, the
quasar has a compact ($<0.1$ arcsec) inverted-spectrum core, which
strongly suggested the presence of a ``central engine'' and encouraged
us to make higher resolution observations with the Multi-Element Radio
Linked Interferometer (MERLIN) and the Very Long Baseline Array
(VLBA).  However, the compact radio emission could also have been
produced by free-free absorption of a compact starburst (Sopp and
Alexander 1991b); even radio variability, present in some radio-quiet
quasars (c.f., Barvainis et al.\ 1996), could be explained by an
ensemble of radio supernovae.

In this letter, we first describe details of our observations with
the VLBA, MERLIN, and the VLA. Following this, we demonstrate that
the compact core detected by the VLBA precludes star formation
as the origin of this radio emission. We then discuss the emission
on scales of hundreds of parsecs shown on the MERLIN map, and
interpret these features in the light of Bridle and Perley's (1984)
criteria for jet identification.

\section{Observations}

We observed E1821+643 for 8 hours on 1996 February 14 using the
10--antenna VLBA (see Napier 1995) of the US
National Radio Astronomy Observatory.  Images in total intensity at 4.9
and 8.4 GHz are shown in Figure 1.  The quasar was observed with the
VLBA in a phase-referenced mode (Beasley and Conway 1995), i.e.,
frequent observations of an adjacent bright source (J1828+64) were
made to provide phase corrections for the interferometer array. The
pointing center was 18$^{\rm h}$ 21$^{\rm m}$ 57.214$^{\rm s}$
64$^{\circ}$ 20$^{\prime}$ 36.231$^{\prime\prime}$ (J2000.0).  
On-source times were 57.6 and 86.4 minutes for the 4.9 and 8.4 GHz
observations respectively.  The rms noise on the two maps
respectively in mJy/beam are 0.12 and 0.079.  The data were processed
using the VLBA correlator, which generated data with four 8.4 MHz
continuum channels in the four Stokes parameters.  Observations of
1928+738 were interspersed throughout the observation to allow
calibration of the polarization response of the receivers.  Our MERLIN
observations were made at 1.7 GHz for 10 hours on 1996 February 8
using the 8 antennas of array including the Lovell telescope.  Phase
referencing was again employed, using 1827+645 as the phase calibrator
and 0552+398 as a point source calibrator.  We concatenated these data
with one of the intermediate frequencies (differing from our chosen
MERLIN frequency by only 7 MHz) from a 40-min observation with the VLA
in BnA array on 1995 September 14.  Synthesis imaging of all data was
performed using the NRAO AIPS system.

\section{Results}

The VLBA maps (Fig.\ 1) show that at 4.9 and 8.4 GHz, the emission on
milli-arcsecond scales is compact: the {\sc fwhm} of the synthesized
beams are $2.4 \times 1.4$ and $1.6 \times 1.0$ milli-arcseconds at
4.9 and 8.4 GHz.  An upper limit to the deconvolved size of the radio
core at 8.4 GHz is $0.3 \times 0.3$ milli-arcseconds, which
corresponds\footnote{One milli-arcsecond corresponds to 5.6\,pc at the
redshift of this quasar, if we assume that ${\em H}_{\circ}~=~50~{\rm
km~s^{-1}~Mpc^{-1}}$, ${\em q_{\circ}}~=~0.5$ and $\Lambda~=~0$.}  to
a physical area of 2.8 pc$^2$.  There are signs that the emission is
very slightly resolved, particularly at 4.9 GHz.

Measurements of peak intensity were made by fitting a Gaussian to each
point source.  The peak fluxes measured in this way, at 4.9 and 8.4
GHz respectively, are $8.6 \pm 0.2$ and $11.9 \pm
0.2$\,mJy/beam.  We
have found with the VLA that at 8.4 GHz the flux density of the radio
core was constant over nearly three years (on 1995 September 14 the
peak in flux density at 8.4 GHz was $12.80 \pm 0.03$ \,mJy/beam while
on 1992 December 15 the peak (at the same resolution) was $12.79 \pm
0.05$\,mJy/beam).  No polarization was detected with the VLBA at 8.4
GHz, down to a 4$\sigma$ limit of 4.2\%.

Using the {\sc fwhm} of the synthesised beam from a uniformly-weighted
map made at 8.4 GHz, we calculated a lower limit to the brightness
temperature (corrected for redshift) of $2.2 \times 10^{8}$\,K.
However, this lower limit can be raised by considering that the
angular size of the emitting region must be considerably smaller than
the point-spread function of the image if the emission is to appear
unresolved.  Use of the (upper limit to the) deconvolved size (above)
gives a lower limit to the brightness temperature of $1.4 \times
10^{9}\ {\rm K}$.

The maps at 1.7 GHz made from MERLIN and VLA data show a number of
components at a resolution of $160 \times 120$ milli-arcseconds.
There is diffuse emission in the vicinity of the core and a second
compact component in the region of the core is found, together with
slightly extended components which seem to follow a curved path
(roughly following features `C' and `D' on our VLA map [Fig.\ 1])
towards knot `E' (involving a change in position angle of $\approx$ 80
degrees). Knot `E' itself is resolved into two components with the
MERLIN data, which lie on a continuation of this curved path.

\section{The origin of the compact radio emission}

We now examine whether radio supernovae or supernova remnants in
star-forming regions can plausibly be retained as the explanation for
the compact radio emission in this radio-quiet quasar. 
Although individual supernovae have brightness temperatures
higher than the brightness temperature of E1821+643, the most luminous
known radio supernova, 1986J (Rupen et al.\ 1987), had a peak
luminosity at 5 GHz of only $\sim 10^{21}\ {\rm W Hz^{-1}}$, so
roughly one thousand of these would be needed to power E1821+643 at 5
GHz. Since the typical lifetime of such a supernova event is $\sim
1$\,yr, this implies a supernova rate $\nu_{\rm SN} \sim 1000$
yr$^{-1}$. Such rates are in line with those required to power the
most luminous RQQs in the starburst scenario (Terlevich
1990). However, to explain the compact radio emission from E1821+643
they must be localized within a few cubic pc, corresponding to a
density $10^7$ times higher than observed in M82 (Muxlow et al.\
1994), and higher than in the starburst model of Terlevich \&
Boyle (1993) by a similar factor.  Although it is possible that in
the dense central region of the nucleus the radio luminosity of
individual supernovae could be substantially enhanced, any reduction
in $\nu_{\rm SN}$ below $\sim 100$ yr$^{-1}$ would be likely to result
in detectable variability on a timescale $\sim 1$ yr. If the
supernovae occur at random, one would expect $\sim 10\%$ variability
for $\nu_{\rm SN} =100$ yr$^{-1}$, higher than the observed limit at
8.4 GHz.  If alternatively we try to explain the radio emission in terms
of supernova remnants, it becomes very difficult to explain the high
brightness temperature observed, since the typical value for such
sources is nearer $10^{4}$\ K (Muxlow et al.\ 1994).

\section{The nature of the core-jet structure}

Of the two compact components on the MERLIN map which are located near
the optical position of the quasar, the more south-westerly of the two
is in good agreement ($10 \pm 15$ mas) with the position of the
compact emission seen on the VLBA image and we therefore identify it
with the flat-spectrum core.  The error in the MERLIN-VLBA
registration is at present dominated by the uncertainty in the MERLIN
phase calibrator position.  The absence of polarization of this
feature found by the VLBA is also consistent with it being a core.

The spectral index of the core was calculated by making a MERLIN map
with a resolution the same (0.17$^{\prime\prime}$) as that of the 15
GHz map described in Blundell \& Lacy (1995).  Using the convention
that $S_{\nu} \propto \nu^{-\alpha}$ (where $S$ is the flux density at
frequency $\nu$) we obtained a spectral index $\alpha^{15}_{1.7} =
-0.83 \pm 0.06$, i.e., as in our VLA study (Blundell and Lacy 1995)
the core spectrum is found to be inverted, though not quite as
steeply.  The second component to the north-east, possibly a jet-knot,
is not detected with our 15 GHz data --- giving a lower limit to the
spectral index of 0.4.

We contend that the feature to the south of the core on the MERLIN map
is a jet.  It appears to satisfy the criterion of Bridle and Perley
(1984) that the jet is at least four times as long as it is wide
(which is true for the knots which follow a curved path from the core
south to knot `E'). Moreover, these features satisfy the other
criteria of Bridle and Perley, namely that they are separable at high
resolution from other extended structure, and they are aligned with
the core where closest: the line joining the two components
closest to the core on either side of it passes through the core.

The curvature of the jet to the south may be consistent with a
scenario in which the jet axis precesses causing the jets to follow
approximately helical paths.  Such a scenario would
also explain the misalignment of the jets and the overall linear
structure of low surface brightness emission (Papadopoulos et al,\
1995) (this misalignment is exaggerated if the quasar is at a small
angle to the line of sight and the jets slightly relativistic). We
will describe our investigations of these and other possibilities,
together with results from low frequency observations with the compact
VLA arrays, in a subsequent paper.

\section{Conclusions} 

We now summarize the evidence that a jet-producing `central engine'
powers this radio source as follows: i) the emission is compact on
similar physical scales to those seen in RLQs (see e.g., Zensus 1994),
ii) the brightness temperature is $\stackrel{>}{_{\sim}}10^9 {\rm K}$
iii) on our MERLIN+VLA maps we see jet-like features on scales of 100
-- 1000 pc, iv) the core luminosity at 5 GHz is $\sim 10^{23}\ {\rm W\,
Hz^{-1} sr^{-1}}$ and arises from a region smaller than a few cubic
pc. Most if not all of the radio emission from this RQQ, just as in
the radio-loud population, is thus powered by a central engine,
probably involving a black hole, rather than star-formation.  To
confirm that the behavior seen in E1821+643 is typical of the
radio-quiet quasar population, we are pursuing a program of VLBI
imaging of a wider sample of RQQs.

\acknowledgments

The VLBA and VLA are facilities of the U.S.\ National Radio Astronomy
Observatory, which is operated by Associated Universities, Inc., under
a co-operative agreement with the US National Science Foundation.
MERLIN is a U.K.\ national facility operated by the University of
Manchester on behalf of PPARC; we are grateful to the Director for the
award of discretionary time on MERLIN.  We are very grateful to Dr
Michael Rupen for helpful discussions, and to Dr Steve Rawlings for
useful comments on the manuscript.

\pagebreak

\pagebreak

\parbox{150mm}{Fig.\ 1 The contour levels in both VLBA maps (which are
both naturally-weighted) are logarithmic with ratio 2; the lowest
levels in mJy/beam are: for the 5\,GHz map, 0.3 and for the 8.4 GHz
map, 0.2.  Where a grey-scale is plotted the units are mJy/beam.  The
VLA image (from Blundell and Lacy [1995]) has its lowest contour at
0.048 mJy/beam; the contours are spaced by factors of $\sqrt{2}$. The
circular beam has {\sc fwhm} $0.3^{\prime\prime}$; co-ordinates are
given in B1950.0 on the VLA map.  The MERLIN map has lowest contour of
0.26 mJy/beam; contours are separated by factors of $\sqrt{2}$. 
The cross on the MERLIN map shows the position of the VLBA core; the total 
extent represents $25 \times $ the uncertainty in the relative position. }

\pagebreak

\hspace*{0cm}
\vspace*{2.5cm}
\begin{picture}(100,300)(0,0)
\put(-75,-475){\includegraphics{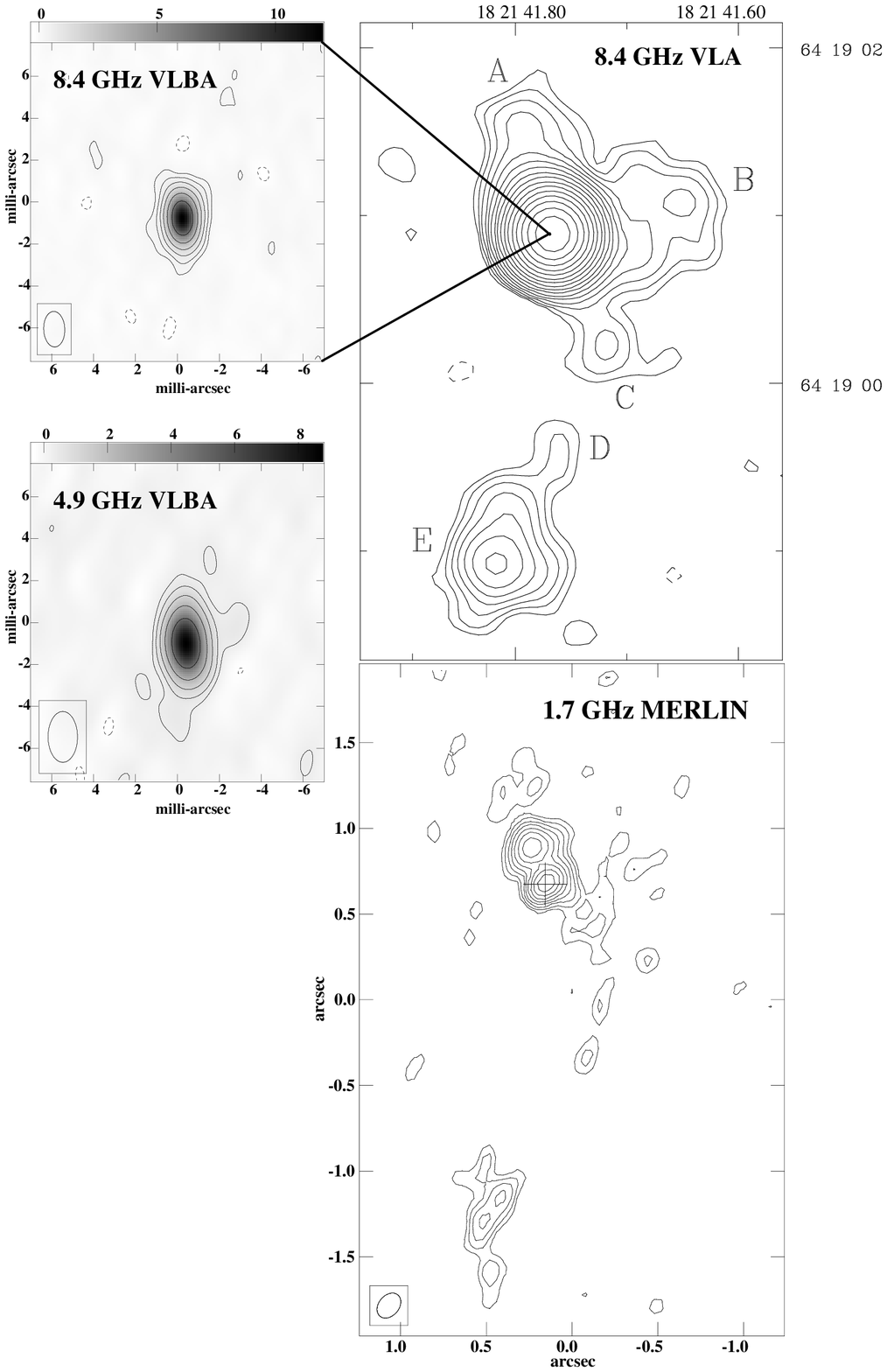}}
\end{picture}

\end{document}